\documentclass[11pt]{article}
\usepackage[utf8]{inputenc}
\usepackage[left = 2.5cm, right = 2.5cm, top = 3cm, bottom = 3cm]{geometry} 
\usepackage{setspace}
\usepackage{lineno}
\usepackage{booktabs, amsmath}
\usepackage[dvipsnames]{xcolor}
\usepackage{sidecap}
\usepackage[hidelinks]{hyperref}
\usepackage{nicefrac}
\usepackage{enumitem} % control enumerate labels
\usepackage{fancyhdr}
\usepackage{tabularx}
\usepackage{amssymb}
\usepackage{mathtools}
\usepackage{amsthm}
\usepackage{bm}
\usepackage{setspace} %Allows double spacing with the \doublespacing command
\usepackage{multirow} % allows for multicolumn or multirow for tables

\usepackage{blindtext}

\usepackage{natbib}
\usepackage{graphicx}
\usepackage{wrapfig}
\usepackage{float}
\usepackage[labelfont=bf, justification=raggedright]{caption}
\DeclareCaptionFont{tenpt}{\fontsize{10pt}{10pt}\selectfont #1}
\title{
The Hill numbers synthesize and generalize measures of trait polygenicity
}

\author{Maike L.~Morrison\thanks{Santa Fe Institute, Santa Fe, NM USA} \thanks{Corresponding author. Email: maike@santafe.edu}}

\date{\today}

\begin{document}
\maketitle

\begin{abstract}
Characterizing the genetic architecture of phenotypic variation is central to human genetics, yet trait polygenicity---the diversity of genetic variants responsible for a heritable trait's variation in a sample---is rarely quantified formally. O'Connor and Sella (2026) addressed this need by proposing four polygenicity measures that are quasi-arithmetic means and therefore satisfy a set of principles they argue are desirable for polygenicity measures. Here, I show that these four measures are points on a continuum corresponding to the Hill numbers, a family of ecological diversity statistics offering a tunable emphasis on variants with small versus large contributions. I further show that the Hill numbers are the only quasi-arithmetic means suitable for measuring polygenicity because they are the only ones that are ``effective numbers'' of variants. The connection to the Hill numbers clarifies the relationships among O'Connor and Sella's four polygenicity measures: all four are effective numbers of variants, and those that give greater weight to variants with small contributions yield higher estimates of polygenicity. It also connects polygenicity to results in the ecological diversity literature, including the replication principle, a criterion for how such measures should behave when independent sets of variants are combined. Together, these results offer a unified framework for interpreting, comparing, and selecting measures of trait polygenicity. 
\end{abstract}

\noindent \textbf{Introduction.} Understanding the genomic basis of phenotypic variation is key to many questions in evolutionary biology and human genetics. While some traits are nearly Mendelian or monogenic, with variation determined almost entirely by a single genetic locus \citep{brinkman2006human}, genome-wide association studies (GWAS) have revealed that the vast majority of traits are polygenic, with variances shaped by a diverse combination of loci, or variants, across the genome \citep{sella2019thinking}. 
The contribution of a variant to the genetic variation of a trait depends both on its effect size and its frequency in the population, generating a wide range of relative contributions across variants \citep{witte2014contribution}. 
The extent to which a trait is polygenic is often discussed but rarely explicitly defined, making it difficult to compare polygenicity across traits. A rigorous measure of polygenicity would need to account for not only the number of variants contributing to the genetic variation of a trait, but also for the relative contribution of each variant to this variation. 

In recent work, \cite{oconnor2026} propose using quasi-arithmetic means to measure polygenicity, since such means satisfy mathematical properties that they argue are desirable for a polygenicity measure. They focus on four particular quasi-arithmetic means as polygenicity measures, demonstrating that they summarize how many variants contribute to the genetic variation of a trait while accounting for each variant's relative contribution. They develop a framework to estimate these polygenicity measures using GWAS summary statistics for 36 complex traits, concluding that most traits are either near-Mendelian or highly polygenic, with few exhibiting intermediate polygenicity. 

At the core of their four measures are different functional forms that appear unrelated, including Shannon entropy and the raw count of variants. In this letter, I show that their measures are closely linked: they are instances of a family of ecological diversity statistics known as the Hill numbers \citep{hill1973diversity}. In the present application, Hill numbers quantify the ``effective number'' of variants contributing to a trait, with a continuous parameter controlling the emphasis placed on ``major'' variants with large relative contributions versus ``minor'' variants with small contributions. The polygenicity measures of \cite{oconnor2026}, despite their apparent differences, are four points along this continuum. Rather than interpreting differences between polygenicity measures in terms of inequalities among their four discrete means \citep[p.~1350]{oconnor2026}, we can recognize a continuum of measures, with those prioritizing minor variants yielding higher polygenicity estimates than those emphasizing major variants. Polygenicity measures that emphasize major variants are more conservative, as they downweight the many minor variants with minuscule contributions. 

Finally, I offer two arguments that the Hill numbers are the most natural measures of polygenicity. First, a polygenicity measure under the definition of \cite{oconnor2026} is only guaranteed to be an effective number of variants if it is mathematically homogeneous \citep{leinster2021entropy}, a property defined below. Requiring that polygenicity measures be effective numbers---and therefore homogeneous---narrows the space of admissible measures from all quasi-arithmetic means to just the Hill numbers. Second, prior work has established the Hill numbers as the only diversity measures satisfying the ``replication principle'' and a set of five further conditions \citep{leinster2021entropy}. I show that the replication principle specifies how polygenicity measures should behave when variants are stratified into non-overlapping subsets. Whereas the first argument selects the Hill numbers from within the family of measures proposed by \cite{oconnor2026}, the second selects them from among all diversity measures without restriction to quasi-arithmetic means. These two independent sets of criteria for polygenicity measures converge on the same family of functions: the Hill numbers.

\vspace{6pt} \noindent \textbf{Definition \& interpretation.}
The Hill numbers, or ``Hill polygenicities,'' depend on two parameters: the vector $\mathbf{p}=(p_1, p_2, \ldots, p_n)$, where $p_i$ is the relative contribution of variant $i$ to the genetic variance of a trait in a sample ($\mathbf{p}$ is called ``$D$'' in \cite{oconnor2026}), and a continuous parameter (or ``order'') $a{\ge}0$ used to tune the emphasis on major versus minor variants (large versus small $p_i$). The Hill polygenicity of order $a$ is defined as:
\begin{equation}
N_a(\mathbf{p})=\left({p_1}^a+{p_2}^a+\ldots +{p_n}^a\right)^{1/(1-a)}.
    \label{eq:hill}
\end{equation}
Each ${p_i}^a$ term shrinks as the exponent $a$ increases because $p_i{<}1$, with smaller contributions shrinking faster than larger contributions \citep{hill1973diversity}. The emphasis placed on minor variants is therefore greatest when $a$ is small, and the emphasis placed on major variants increases as $a$ increases.

The Hill polygenicities can be interpreted as ``effective numbers of variants''---if the Hill polygenicity (for a given order, $a$) of a trait is equal to 1,000, then the trait is as polygenic as a hypothetical trait with 1,000 equally contributing variants. This property is somewhat obscured in \cite{oconnor2026}, where just one of the four proposed polygenicity measures is named the ``effective polygenicity,'' when in fact their discrete measures (their equations 1, 2, 3, and 7) all share this interpretation. 

\vspace{6pt} \noindent \textbf{The four polygenicity measures of O'Connor and Sella are instances of the Hill polygenicities.}
The  Hill polygenicities of order 0, 1, 2, and infinity are equivalent to the four discrete polygenicity measures of \citeauthor{oconnor2026}.
\begin{itemize}
    \item \underline{$a=0$} gives $N_0(\mathbf{p})=(1+1+\ldots+1)^{1/(1-0)}=n$, which is the number of variants with nonzero contribution to the genetic variation of a trait. This expression is equivalent to the arithmetic measure of \cite{oconnor2026} (their equation 1).  This quantity places maximal emphasis on minor variants, since each adds 1 to the value of $N_0(\mathbf{p})$ irrespective of its contribution to genetic variation. 
    \item \underline{$a\to1$} gives $N_1(\mathbf{p})=\exp \left( -\sum_{i=1}^np_i\ln p_i \right)$, which is the exponentiated Shannon entropy. This expression is equivalent to the geometric measure of \cite{oconnor2026} (their equation 3, also called the ``entropy polygenicity'').
    \item \underline{$a=2$} gives $N_2(\mathbf{p})=1\big/\left({p_1}^2+{p_2}^2+\ldots +{p_n}^2\right)$, which is the inverse Simpson index. This expression is equivalent to the harmonic measure of \cite{oconnor2026} (their equation 2, also called the ``effective polygenicity''). 
    \item \underline{$a\to\infty$} gives $N_{\infty}(\mathbf{p})=1 /\max_i(p_i)$, which is the reciprocal of the relative contribution of the largest-contribution variant. This expression is equivalent to equation 7 of \cite{oconnor2026}, which they approximate with a measure called ``softmax'' (their equation 8).  $N_{\infty}$ places the maximal possible emphasis on major variants, since it is a function of only the largest-contribution variant, $\max_i(p_i)$.
\end{itemize}

To close this section, I show the correspondence between the general measure of \cite{oconnor2026} and the Hill numbers.
\cite{oconnor2026} frame their four polygenicity measures as weighted quasi-arithmetic means of the effective polygenicities of individual variants ($\nicefrac{1}{p_i}$), with each measure associated with a different function $f$:
\begin{equation}
    \label{eq:power_mean} M_f(\mathbf{p})=f^{-1}\left(\sum_{i=1}^np_i f(\nicefrac{1}{p_i}) \right)
\end{equation}
(their equation 4). Their first three measures are the three Pythagorean means,  equivalent to the quasi-arithmetic means associated with $f(x)=x$ (arithmetic), $f(x)=\ln x$ (geometric), and $f(x)=\nicefrac{1}{x}$ (harmonic). They approximate their fourth measure, the maximum effective polygenicity across variants, using a quasi-arithmetic mean with $f(x)=\exp(-x)$. 

Their four measures can instead be viewed within the broader family of power means, the quasi-arithmetic means associated with $f(x)=x^b$ \citep{hardy1952inequalities}: 
\begin{equation} \label{eq:hill_power}
    M_b(\mathbf{p})=
    \left[\sum_{i=1}^np_i (\nicefrac{1}{p_i})^b \right]^{1/b}.
\end{equation}
Equation \ref{eq:hill_power} is equivalent to our previous formulation of the Hill numbers, $N_a(\mathbf{p})$ (Equation \ref{eq:hill}), when $b=1-a$ \citep[ch.~4.2]{leinster2021entropy}. 
Framing the measures of \cite{oconnor2026} as power means unifies their four functional definitions within a single continuous function, $f(x)=x^b$ (or, equivalently, $f(x)=x^{1-a}$), that can be used in their equation 15 to estimate polygenicity from GWAS summary statistics. It also makes explicit the precise mathematical parallel between their measures and the Hill numbers. Here, this correspondence serves a specific goal: I will use properties of power means to demonstrate that the Hill numbers are the only quasi-arithmetic means suitable for polygenicity measurement.  

\vspace{6pt} \noindent \textbf{Principles specifying the Hill numbers.}
If the quasi-arithmetic means are defined for many functions $f$, why should we constrain possible polygenicity measures to the power means, which use $f(x)=x^b$? The answer lies in the notion of first-degree homogeneity, which specifies that rescaling a function's inputs by a constant is equivalent to rescaling its output by the same constant. Intuitively, this means that a function must be homogeneous in order for its inputs and outputs to have the same units \citep[p.~240]{leinster2021entropy}. In this setting, our function's output is a measure of polygenicity, and its inputs are the effective polygenicities ($\nicefrac{1}{p_i}$), whose units are a number of variants, since the reciprocal of a proportion is a number. If we want our polygenicity measure to also be a number of variants, we must measure polygenicity with a homogeneous function. Because power means are the only homogeneous quasi-arithmetic means, the Hill numbers (Equation \ref{eq:hill_power}) are the only polygenicity measures among all possible quasi-arithmetic means that yield an effective number of variants. 

An alternative approach is to derive the Hill numbers from first principles, without presupposing that polygenicity has to be measured by a quasi-arithmetic mean. \citeauthor{oconnor2026} asked how the total polygenicity of a set of variants should relate to the polygenicities of its non-overlapping subsets (e.g., coding versus non-coding) in order to highlight a helpful property of their entropy polygenicity measure, $N_1(\mathbf{p})$. The ecological diversity statistics literature answers a similar question differently, with what is known as the ``replication principle''  \citep{Chao2014, leinster2021entropy}. Rephrased in the language of polygenicity, it states that if a trait is shaped by $K$ non-overlapping subsets of genetic variants, each set contributing equally to the trait's genetic variance and each with polygenicity $P$, then the overall polygenicity of the trait should be $K\cdot P$. Together with five further conditions, the replication principle exactly specifies the Hill numbers from among all possible diversity measures (Theorem 7.4.3 of \cite{leinster2021entropy}), without restriction to only quasi-arithmetic means. These six axioms offer an alternative to the set proposed by \cite{oconnor2026}. 

The replication principle prescribes intuitive behavior for polygenicity under partition, making it a natural criterion for polygenicity measures. For example, suppose two non-overlapping sets of variants each contribute half of a trait's genetic variance and each set has polygenicity $P$. The replication principle requires that the total polygenicity of the trait be $2P$. While the Hill numbers satisfy the replication principle, other seemingly useful metrics of polygenicity do not. For example, it could seem sensible to measure polygenicity of different traits by comparing their values of Shannon entropy. Although Shannon entropy is often used to quantify the evenness of a distribution, it does not satisfy the replication principle and in this example would assign the trait a total polygenicity less than $2P$. (Note that $N_1(\mathbf{p})$, the exponential of Shannon entropy, does obey the replication principle.) Similarly, the softmax polygenicity of \cite{oconnor2026}, while potentially a useful approximation, does not satisfy the replication principle, equaling $2P$ in the above example only if the contributions within each set are perfectly even and otherwise yielding a total polygenicity less than $2P$. That two otherwise reasonable measures fail the replication principle suggests it is not a trivial requirement, but a substantive constraint that is notably satisfied by the Hill numbers.

\vspace{6pt} \noindent \textbf{Examples.}
To illustrate the interpretation of the Hill polygenicities, consider three  scenarios for the relative contributions of $n{=}10{,}000$ genetic variants to the variation of a trait. First, suppose all 10,000 variants contribute 0.01\%:
$$\mathbf{p_A}=\big(\underbrace{10^{-4},  10^{-4}, \ldots, 10^{-4}}_{10{,}000}\big).$$
Because the contributions are perfectly even across the 10,000 variants, all orders of the Hill polygenicity equal 10,000, and so the four discrete measures of \citeauthor{oconnor2026} are identical: $N_0(\mathbf{p_A})=N_1(\mathbf{p_A})=N_2(\mathbf{p_A})=N_\infty(\mathbf{p_A})=10{,}000$. In general, if there are $n$ variants with identical contributions to a trait, the trait will have a Hill polygenicity of $n$ irrespective of how major versus minor variants are weighted.

Second, suppose there is a more complex distribution of contributions across variants: 200 variants contribute 0.05\%, 7,800 variants contribute 0.01\%, and the remaining 2,000 variants contribute 0.006\%: 
$$\mathbf{p_B}=\big(\underbrace{5\times10^{-4},  \ldots, 5\times10^{-4}}_{200},
\underbrace{10^{-4},  \ldots, 10^{-4}}_{7{,}800}, 
\underbrace{6\times10^{-5},  \ldots, 6\times10^{-5}}_{2{,}000}\big).$$ 
When applied to this scenario, the four discrete measures are: $N_0(\mathbf{p_B})=10{,}000$,  $N_1(\mathbf{p_B})\approx 9{,}051$, $N_2(\mathbf{p_B})\approx 7{,}396$, and $N_\infty(\mathbf{p_B})=\nicefrac{1}{0.0005}=2{,}000$. The $a{=}0$ order captures that there are 10,000 variants contributing in some way to the trait. The infinity order indicates that there are effectively 2,000 variants contributing to the trait if only the contribution of the most dominant one is considered. The gradual taper of Hill polygenicity between $a=0$ and $a=\infty$ captures that there are many variants with  substantial contributions, but their contributions are not completely even.

Finally, suppose there is a single major variant contributing 99\% of the variation, and 9,999 minor ones dividing the remainder: 
$$\mathbf{p_C}=\big(0.99, \underbrace{\nicefrac{0.01}{9999}, \ldots, \nicefrac{0.01}{9999}}_{9999}\big).$$
Again, $N_0(\mathbf{p_C})=10{,}000$, but $N_1(\mathbf{p_C})\approx 1.16$, $N_2(\mathbf{p_C})\approx 1.02$, and $N_\infty(\mathbf{p_C})=\nicefrac{1}{0.99}\approx 1.01$. The fact that $N_a(\mathbf{p_C})\approx 1$ when $a$ is not close to 0 reflects that, when variant contributions are considered, the trait is effectively monogenic.

Collectively, these examples highlight how the Hill numbers unify the polygenicity measures proposed by \cite{oconnor2026} and allow them to be interpreted through the lens of varying emphasis on major versus minor variants.

\begin{figure}[t]
  \begin{center}
    \includegraphics[width=0.7\textwidth]{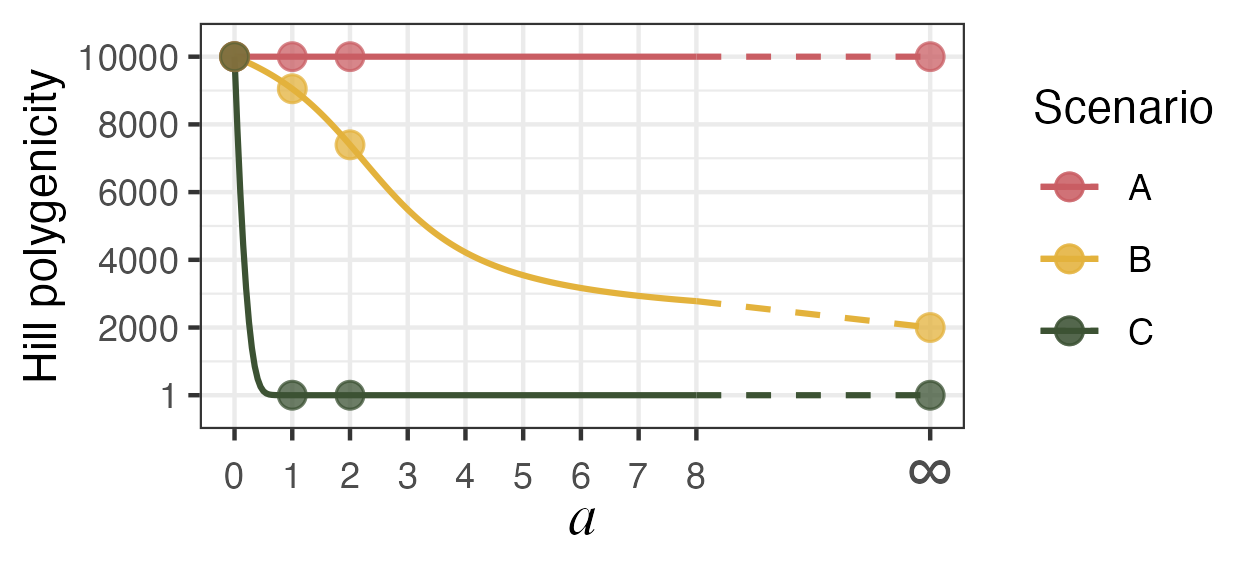}
  \end{center}
  \vspace{-24pt}
  \caption{\textbf{Polygenicity profiles.} Hill polygenicity $N_a(\mathbf{p})$ as a function of $a$ for scenarios A ($\mathbf{p_A}$), B ($\mathbf{p_B}$), and C ($\mathbf{p_C}$). Larger $a$ places greater emphasis on major variants. Points mark the four measures of \cite{oconnor2026}: $a=0,1,2,\infty$. The dashed line marks the approach from $a=8$ to the infinite limit, $N_\infty(\mathbf{p})$.}  \label{fig:profile}
  \vspace{-12pt}
\end{figure}

\vspace{6pt} \noindent \textbf{Continuous polygenicity profiles.}
The Hill polygenicities are not limited to the four measures that were the focus of \cite{oconnor2026}, but are defined for any $a{\ge}0$. This property enables researchers to plot the ``polygenicity profile'' of a trait, with the order $a$ on the x-axis and the corresponding Hill polygenicity of the trait, $N_a(\mathbf{p})$, on the y-axis (e.g.,  Figure \ref{fig:profile}). The polygenicity profile of a trait whose $n$ variants have equal contributions (e.g., scenario A) will be a horizontal line at $N_a(\mathbf{p})=n$. All other genetic architectures have monotonically declining polygenicity profiles, since $N_a(\mathbf{p})$ is non-increasing in $a$ \citep[p.~119]{leinster2021entropy}.  A trait driven by $n$ variants with a mixture of large and small contributions will have a y-intercept of $n$, but will decay to an asymptote equal to the reciprocal of the contribution of the largest-contribution variant (i.e., $N_{\infty}(\mathbf{p})=\nicefrac{1}{\max_i(p_i)}$), with traits more dominated by fewer variants showing a more rapid rate of decay (e.g., scenario C). 

Such profiles can improve intuition regarding the expected relationships between polygenicity measures (i.e., different values of $a$). Indeed, whereas  \cite{oconnor2026} use inequalities among Pythagorean means to explain why some polygenicity measures exceed others, polygenicity profiles demonstrate that a measure of polygenicity will be higher when it assigns greater importance to minor variants (i.e., small $a$), and lower when the measure downweights minor variants  (i.e., large $a$)   \citep[p.~119]{leinster2021entropy}. This interpretation also reframes a problem raised in the introduction of \cite{oconnor2026} regarding variants with arbitrarily small contributions: whereas the Hill polygenicity of order $a=0$ counts all variants equally, small contributions are ignored when $a$ is large. Finally, polygenicity profiles can clarify differences between traits, complementing plots of just a few selected measures (e.g., Figure 3A of \cite{oconnor2026}).

\vspace{6pt} \noindent \textbf{Conclusions.} 
\cite{oconnor2026} made a substantial advance in polygenicity measurement by deriving principled measures of trait polygenicity and estimating these measures from GWAS summary statistics. Here, I have shown that their four focal measures correspond to the 0, 1, 2, and infinity orders of the Hill numbers \citep{hill1973diversity}, a family of ecological diversity statistics. Viewing their measures as Hill numbers reveals them as four points along a continuum of emphasis on major versus minor variants and explains the inequalities among them, since measures that emphasize minor variants yield higher polygenicity estimates. This perspective also  provides a natural way to address the influence of variants with contributions that are vanishingly small. Moreover, the Hill numbers extend the four focal measures of \citeauthor{oconnor2026} into a continuous family, enabling the visualization of ``polygenicity profiles'' that facilitate comparisons across traits. The Hill numbers can be directly integrated into the estimation framework of \cite{oconnor2026} by using $f(x)=x^{1-a}$ in their equation 15. 

Two distinct arguments identify the Hill numbers as natural measures of polygenicity. First, among the quasi-arithmetic means proposed by \cite{oconnor2026}, only the Hill numbers are homogeneous, making them the only means that measure polygenicity as an effective number of variants. Second, without restriction to quasi-arithmetic means, the Hill numbers are the only diversity measures satisfying both the replication principle and five further conditions specified by \cite{leinster2021entropy}. This convergence of two different sets of criteria offers complementary justification for measuring polygenicity with the Hill numbers, and broadens the axiomatic basis for polygenicity beyond the quasi-arithmetic means. Together, these interpretations, extensions, and principles provide a unified framework for measuring and understanding trait polygenicity.

\subsection{Acknowledgments} MLM thanks Graham Coop and Schuyler Nicholson for feedback and constructive conversations, and Luke O'Connor and Guy Sella for helpful correspondence. Funding was provided by the National Institutes of Health (R35 GM136290, awarded to GC) and the Santa Fe Institute Omidyar Fellowship (awarded to MLM). 

\bibliography{references}

\end{document}